\documentclass[12pt,a4paper,dvips]{article}
\usepackage{a4p}
\usepackage{cite,mcite}
\usepackage{graphicx}
\usepackage{physics}
\usepackage{eepic}                      
\usepackage{feynmf}
\usepackage{epsfig}                      
\usepackage{l3_title,ifthen}
\usepackage{amsmath}
\usepackage{dcolumn}
\journalname{Phys. Lett. B}
\date{December 3, 2003}
\preprint{2003-080}
\newlength{\capindent}
\newlength{\capwidth}
\newlength{\figwidth}
\newcommand{\icaption}[2][!*!,!]{\hspace*{\capindent}%
  \begin{minipage}{\capwidth}
    \ifthenelse{\equal{#1}{!*!,!}}%
      {\caption{#2}}%
      {\caption[#1]{#2}}
  \end{minipage}}
\newcolumntype{-}{D{-}{~\mbox{--}~}{4}}
\newcolumntype{/}{D{/}{~\pm~}{2}}
\newcolumntype{+}{D{+}{~\pm~}{12}}
\newcolumntype{a}{D{a}{~\pm~}{11}}

 {\end{list}}
\begin{document}
\begin{titlepage}
\title{ Muon-Pair and Tau-Pair Production in \\ Two-Photon Collisions at LEP}

\author{L3 Collaboration}

\begin{abstract}

The QED processes $\ee \rightarrow \ee \mu^+\mu^-$ and $\ee
 \rightarrow \ee \tau^+\tau^-$ are studied with the L3 detector at LEP
 using an untagged data sample collected at centre-of-mass energies
 $161 \GeV \le\ \rts\ \le\ 209 \GeV$. The $\tau$-pairs are
 observed through the associated decay of one $\tau $ into $\e\nu\nu$
 and the other into $\pi\pi\nu$.  The cross sections are measured as a
 function of $\rts$.  For muon pairs, the cross section of
 the $\gamma\gamma \rightarrow \mu^+\mu^-$ process is also measured as a function of the
 two-photon centre-of-mass energy for $ 3 \GeV \le\ W_{\gamma\gamma}\
 \le 40 \GeV $. Good agreement is found between these measurements and
 the $\mathcal{O}(\alpha^4)$ QED expectations. In addition, limits on
 the anomalous magnetic and electric dipole moments of the $\tau$
 lepton are extracted.

\end{abstract}

\submitted

\end{titlepage}

\section{Introduction}

The pair production of charged leptons in two-photon collisions offers a
unique opportunity to test QED to $\mathcal{O}(\alpha^4)$ over a wide
kinematical range. The $\ee \rightarrow \ee \mu^+ \mu^-$ and $\ee
\rightarrow \ee \tau^+ \tau^-$ reactions are studied with the L3
detector\cite{L3} for untagged events, in which the $\e^+$ and $\e^-$,
scattered at small angles, are not required to be observed.

Figure~\ref{fig:feyn} shows the lowest order processes which
contribute to this final state: multiperipheral, bremsstrahlung,
annihilation and conversion, for a total of 12 possible diagrams.  For
untagged events the multiperipheral process dominates the cross
section. The contribution of other processes is below 1\%.

The $\ee \rightarrow \ee \ell^+ \ell^-$ reactions, where $\ell = \e,
\mu$ or $\tau$, were previously studied for untagged two-photon events
at $\ee$ centre-of-mass energy, $\rts$, close to the Z mass
~\cite{emutauL3}. Good agreement was found between the measurements
and the QED expectations.  In this Letter, the production of $\mu$-pairs is studied in the range $161 \GeV \le \rts \le 209
\GeV$ and the production of $\tau$-pairs in the range $189 \GeV \le \rts \le 209 \GeV$.  The individual energies and luminosities are listed in
Table~\ref{tab:selevents}.  As the $\ee \rightarrow \ee \mu^+ \mu^-$
channel benefits from high statistics, the cross section of the
process $\gamma \gamma \rightarrow \mu^+ \mu^-$ is also measured as a
function of the two-photon centre-of-mass energy, $W_{\gamma\gamma}$.
The $\ee \rightarrow \ee \tau^+ \tau^-$ analysis is restricted to the
exclusive final state where $\tau^- \to \e^- \nu_\tau\bar{\nu_\e}$ and
$\tau^+ \to \pi^+\pi^0\bar{\nu_\tau}$\footnote{Charge conjugate
processes are included throughout this Letter.}, which arises from
$9.07\pm0.01\%$
of all $\tau$-pair decays \cite{PDG}.

The process $\ee \rightarrow \ee \tau^+ \tau^-$ is also used to
constrain the anomalous magnetic and electric dipole moments of the
$\tau$ lepton, as proposed in Reference~\citen{Cornet:1996pw}.

\section{Data and Monte Carlo Samples}

The events are mainly accepted by the charged-particle~\cite{trig1}
and the inner track triggers~\cite{trig2}. The former requires at
least two charged particles with a transverse momentum $p_{t}>150
\MeV$, back-to-back within an acoplanarity angle of $\pm 41^{\circ}$.
The latter is based on a neural network, has no requirement on the
acoplanarity angle of the tracks and extends the acceptance from the
polar region $30^{\circ}<\theta<150^{\circ}$ to
$15^{\circ}<\theta<165^{\circ}$.  A fraction of the $\ee \rightarrow
\ee \mu^+ \mu^-$ events is also accepted by the muon trigger and a
fraction of the $\ee \rightarrow \ee \tau^+ \tau^-$ events by the
calorimetric energy trigger~\cite{trig3}.

\begin{figure}[p]
\begin{center}

\begin{fmffile}{ggll}

a)\hspace{1.3cm}\begin{fmfgraph*}(100,75) \fmfpen{thick} \fmfstraight
    \fmfleft{i1,i2}\fmfrightn{o}{4}
    \fmfforce{.5w,.2h}{v1}
    \fmfforce{.5w,.4h}{v2}
    \fmfforce{.5w,.6h}{v3}
    \fmfforce{.5w,.8h}{v4}
    \fmf{fermion}{o1,v1,i1}
    \fmf{fermion}{i2,v4,o4} 
    \fmf{boson}{v1,v2}
    \fmf{boson}{v4,v3}
    \fmf{fermion}{o2,v2,v3,o3}
    \fmfdot{v1,v2,v3,v4}
    \fmflabel{$\mathrm{e}^+$}{i1}
    \fmflabel{$\mathrm{e}^-$}{i2}
    \fmflabel{$\mathrm{e}^+$}{o1}
    \fmflabel{$\tau^+/\mu^+$}{o2}
    \fmflabel{$\tau^-/\mu^-$}{o3}
    \fmflabel{$\mathrm{e}^-$}{o4}
  \end{fmfgraph*} \hspace{2.2cm} b) \hspace{1.3cm}
  \begin{fmfgraph*}(100,75) \fmfpen{thick} \fmfstraight
    \fmfleft{i1,i2}\fmfrightn{o}{4}
    \fmfforce{.4w,.2h}{v1}
    \fmfforce{.4w,.8h}{v2}
    \fmfforce{.6w,.13h}{v3}
    \fmfforce{.8w,.5h}{v4}
    \fmf{fermion}{o1,v3,v1,i1}
    \fmf{fermion}{i2,v2,o4} 
    \fmf{boson}{v1,v2}
    \fmf{boson}{v3,v4}
    \fmf{fermion}{o2,v4,o3}
    \fmfdot{v1,v2,v3,v4}
    \fmflabel{$\mathrm{e}^+$}{i1}
    \fmflabel{$\mathrm{e}^-$}{i2}
    \fmflabel{$\mathrm{e}^+$}{o1}
    \fmflabel{$\tau^+/\mu^+$}{o2}
    \fmflabel{$\tau^-/\mu^-$}{o3}
    \fmflabel{$\mathrm{e}^-$}{o4}
  \end{fmfgraph*} 

\vspace{1.5cm}

c) \hspace{1.3cm} \begin{fmfgraph*}(100,75) \fmfpen{thick} \fmfstraight
    \fmfleft{i1,i2}\fmfrightn{o}{4}
    \fmfforce{.3w,.2h}{v1}
    \fmfforce{.3w,.8h}{v2}
    \fmfforce{.7w,.2h}{v3}
    \fmfforce{.7w,.8h}{v4}
    \fmf{fermion}{i2,v2,v1,i1}
    \fmf{boson}{v1,v3}
    \fmf{boson}{v2,v4}
    \fmf{fermion}{o2,v3,o1}
    \fmf{fermion}{o4,v4,o3}
    \fmfdot{v1,v2,v3,v4}
    \fmflabel{$\mathrm{e}^+$}{i1}
    \fmflabel{$\mathrm{e}^-$}{i2}
    \fmflabel{$\mathrm{e}^+$}{o2}
    \fmflabel{$\tau^+/\mu^+$}{o4}
    \fmflabel{$\tau^-/\mu^-$}{o3}
    \fmflabel{$\mathrm{e}^-$}{o1}
  \end{fmfgraph*} \hspace{2.2cm}  d) \hspace{1.3cm}
  \begin{fmfgraph*}(100,75) \fmfpen{thick} \fmfstraight
    \fmfleft{i1,i2}\fmfrightn{o}{4}
    \fmfforce{.2w,.5h}{v1}
    \fmfforce{.5w,.5h}{v2}
    \fmfforce{.7w,.3h}{v3}
    \fmfforce{.85w,.5h}{v4}
    \fmf{fermion}{i2,v1,i1}
    \fmf{fermion}{o1,v3,v2,o4} 
    \fmf{boson}{v1,v2}
    \fmf{boson}{v3,v4}
    \fmf{fermion}{o2,v4,o3}
    \fmfdot{v1,v2,v3,v4}
    \fmflabel{$\mathrm{e}^+$}{i1}
    \fmflabel{$\mathrm{e}^-$}{i2}
    \fmflabel{$\mathrm{e}^+$}{o1}
    \fmflabel{$\tau^+/\mu^+$}{o2}
    \fmflabel{$\tau^-/\mu^-$}{o3}
    \fmflabel{$\mathrm{e}^-$}{o4}
  \end{fmfgraph*} 

\end{fmffile}

\vspace{0.5cm}
\caption{Feynman graphs at ${\cal O}(\alpha^4)$ of the
  processes $\ee \to \ee \mu^+\mu^-$ and $\ee \to \ee \tautau$:
a) multiperipheral, b) bremsstrahlung, c) conversion and d) annihilation.
}
\label{fig:feyn}
\end{center}
\end{figure}
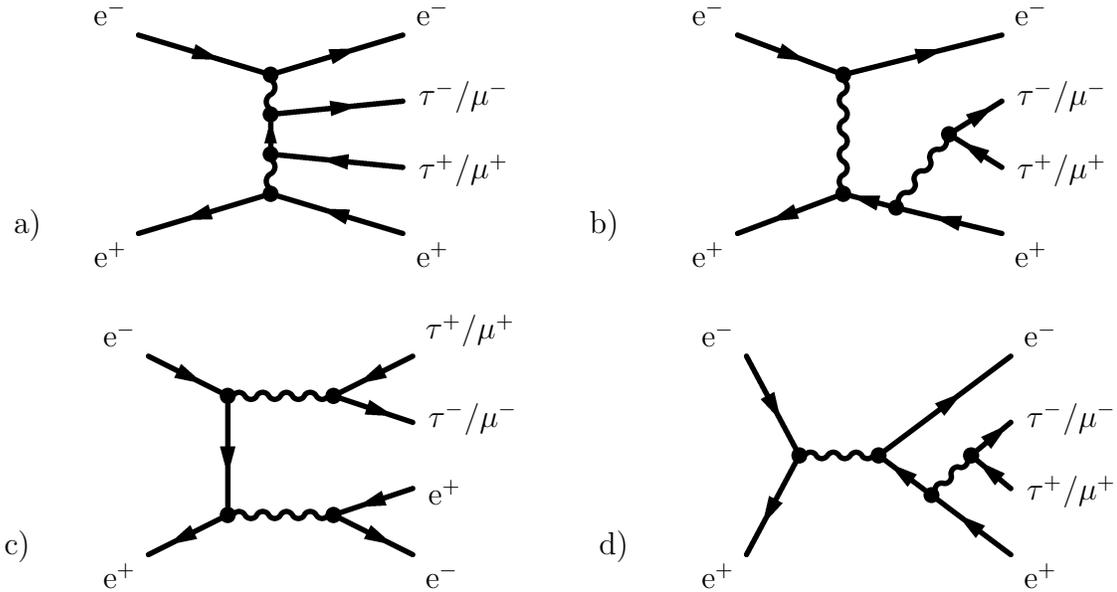

The DIAG36 \cite{BDK} generator is used to calculate at
$\mathcal{O}(\alpha^4)$ the full set of diagrams shown in
Figure~\ref{fig:feyn}.  To obtain the efficiencies of the $\ee
\rightarrow \ee \mu^+ \mu^-$ channel, high statistics samples are
generated in the range $ 3 \GeV \le\ W_{\gamma\gamma}\ \le 40 \GeV $,
for each value of $\rts$.  The $\ee
\rightarrow \ee \tau^+ \tau^-$ events are generated in the full phase
space with the Vermaseren Monte Carlo~\cite{Vermaseren}, which takes
into account only the dominating multiperipheral diagrams, shown in
Figure \ref{fig:feyn}a.
 
For background studies, the following event generators are used:
KORALZ \cite{Koralz} for the $\ee \rightarrow \tau ^+\tau ^-$ and $\ee
\rightarrow \mu^+\mu^-$ processes and LEPWW \cite{lepww} and PYTHIA
\cite{pythia} for W and Z boson pair-production and decays into
leptons, respectively. In the tau-pair analysis, exclusive hadronic
two-photon processes are generated with EGPC~\cite{EGPC} and inclusive
hadron production with PHOJET~\cite{Phojet}.

All generated events are processed through the full L3 detector
simulation based on the GEANT~\cite{GEANT} and GHEISHA~\cite{GEISHA}
programs and are reconstructed following the same procedure as for
the data. Time dependent detector inefficiencies, as monitored during
the data taking, are also included.

\section{Event Selection} 

\subsection{\boldmath $\rm e^+e^- \rightarrow e^+e^- \mu^+ \mu^-$}

The muon pairs are selected using information from the central
tracking chamber (TEC) and the muon spectrometer. The selection
requires:

\begin{itemize}
\item exactly two tracks with at least 12 hits each and opposite
charges, having a distance of closest approach to
the nominal interaction vertex in the plane transverse to the beam
direction smaller than 5 mm;
\item two well reconstructed muons in the muon chambers corresponding
to the charged tracks;
\item a fiducial volume $|\cos \theta_\mu| < 0.8$,  where  $\theta_\mu$
is the angle between the muon and the beam axis;
\item the momentum of the muons between 2.5 \GeV\,and 40 \GeV;
\item muon tracks pointing to the primary vertex, with time-of-flight
consistent with the beam crossing, in order to suppress background
from cosmic rays, hadrons decaying in flight and punch-through
hadrons;
\item a di-muon effective mass, $M_{\mu\mu}$, which measures
$W_{\gamma\gamma}$, between 3 \GeV\,and 40 \GeV.
\end{itemize}

\begin{figure}[p]
\begin{center}
\epsfig{file=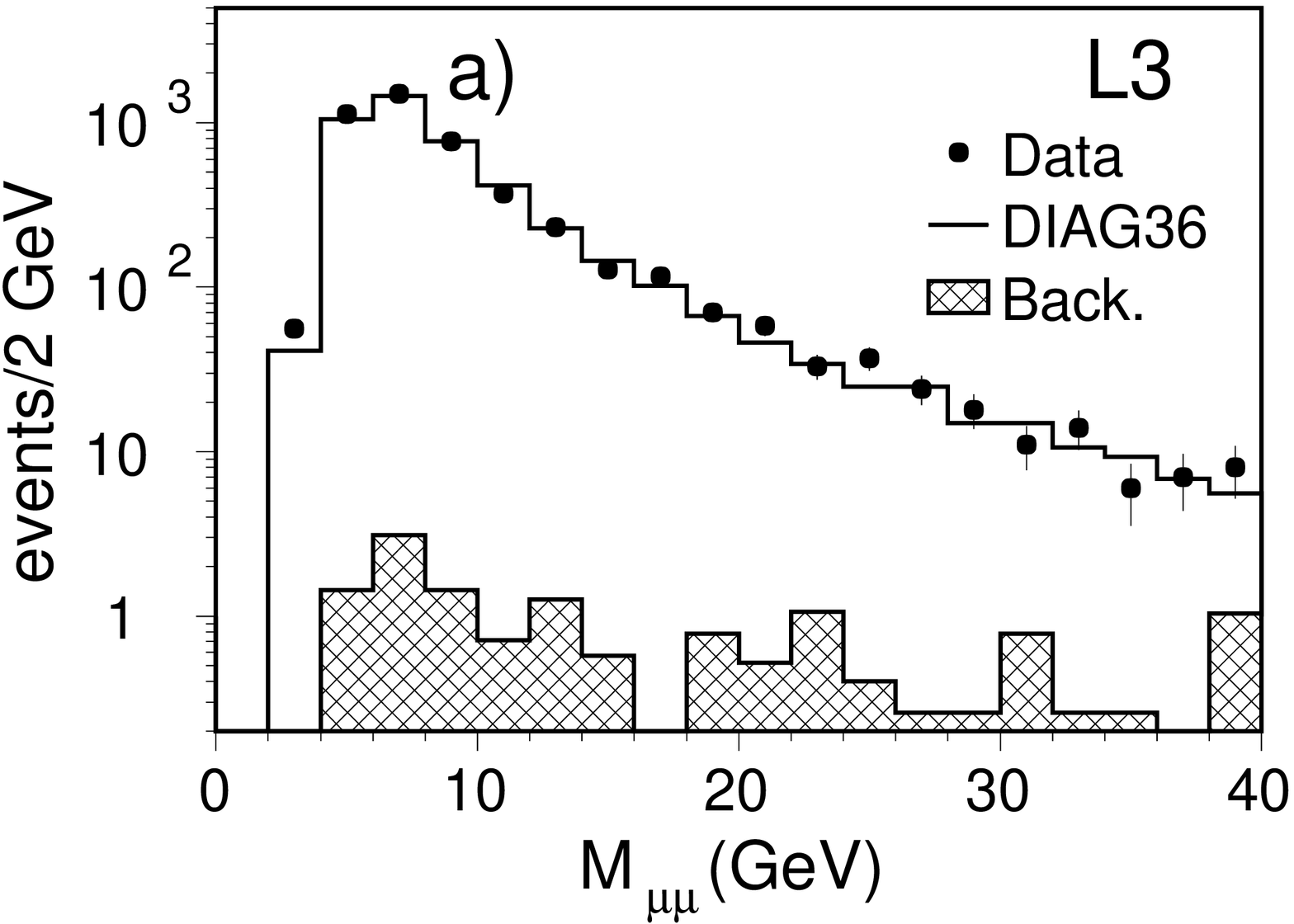,width=140mm,clip=}
\epsfig{file=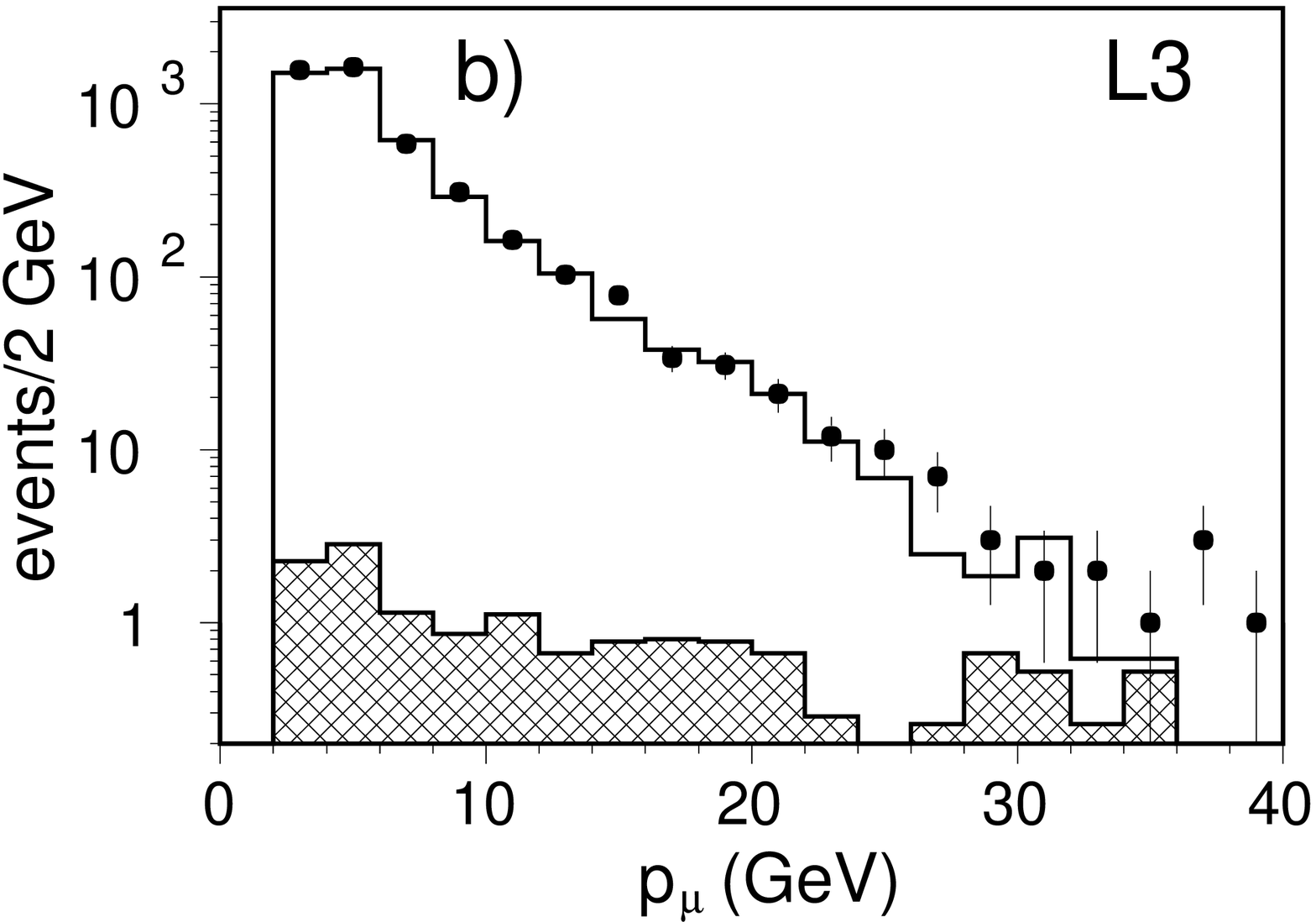,width=140mm,clip=}
\caption{Distributions for selected $\ee \to \ee\mu^+\mu^-$ events of
  a) the di-muon effective mass, $M_{\mu\mu}$, and b) the momentum of
   the most energetic muon, $p_\mu$. The data are compared to the sum
   of the DIAG36 Monte Carlo and of the expected background,
   normalized to the integrated luminosity.}
\label{fig:musel}
\end{center}
\end{figure}

The numbers of events selected at different $\rts$ are shown in
Table~\ref{tab:selevents} together with the selection and trigger
efficiencies. The total background contribution, estimated by Monte
Carlo, is below $ 1\%$, and consists mainly of events from the $\ee
\to \ee\tau^+\tau^-$, $\ee \to \tau^+\tau^-$ and $\ee \to \mu^+\mu^-$
processes and cosmic rays.  The distributions of the di-muon effective
mass and of the momentum of the higher energy muon are presented in
Figure~\ref{fig:musel}a and~\ref{fig:musel}b together with the Monte
Carlo predictions. The expected distributions agree well with the data.
 
\subsection{\boldmath$\ee \rightarrow \ee \tau^+ \tau^-$}

The selection of tau-pairs, through the associated decays $\tau^- \to
\e^-\nu_\tau\bar{\nu_\e}$ and $\tau^+ \to \pi^+\pi^0\bar{\nu_\tau}$, is 
based on information from the TEC and the electromagnetic calorimeter (ECAL). It requires:

\begin{itemize}
\item a total energy in the calorimeters less than 40 \GeV, to exclude
$\ee \to \tau^+\tau^-$ events;
\item exactly two charged tracks with at least 12 hits each and
opposite charges, having a transverse momentum greater than $ 0.3
\GeV$, a distance of closest approach to the
nominal interaction vertex in the plane transverse to the beam
direction smaller than 10 mm and a corresponding ECAL signal;
\item two photons, defined as isolated showers in the ECAL with energy
greater than 100 \MeV\,distributed over at least two crystals. There
must be no track within 150 mrad around the shower direction and the
ratio between the energies deposited in the hadronic and
electromagnetic calorimeters must be less than $ 0.2$.
\end{itemize}

The electron identification for the reaction $\tau^- \rightarrow
\e^-{\nu_{\tau}} \bar{\nu_\e}$ is based on an ECAL cluster, with a
shower shape consistent with that of an electromagnetic particle,
matching with a charged track within \mbox{100\,mrad} in the plane
transverse to the beam direction.  The momentum of the electron
candidate must be greater than $ 600$~\MeV. To achieve high efficiency
and high purity, the electron identification is based on a neural
network~\cite{SNNS} which combines ten variables: the energy in ECAL,
the momentum, the ionisation energy loss in TEC, the ratio of the
transverse energy in ECAL to the transverse momentum in TEC, the
number of crystals in the shower, three inputs describing the shower
shape in ECAL, the corresponding energy in the hadronic calorimeter
and its fraction within a $7^\circ$ cone.  The electron identification
with the neural network has an efficiency of $87.7\pm 0.2$\% with a
purity of $94.7\pm 0.2$\%, as determined from Monte Carlo events.

\begin{figure}[p]
\begin{center}
\epsfig{file=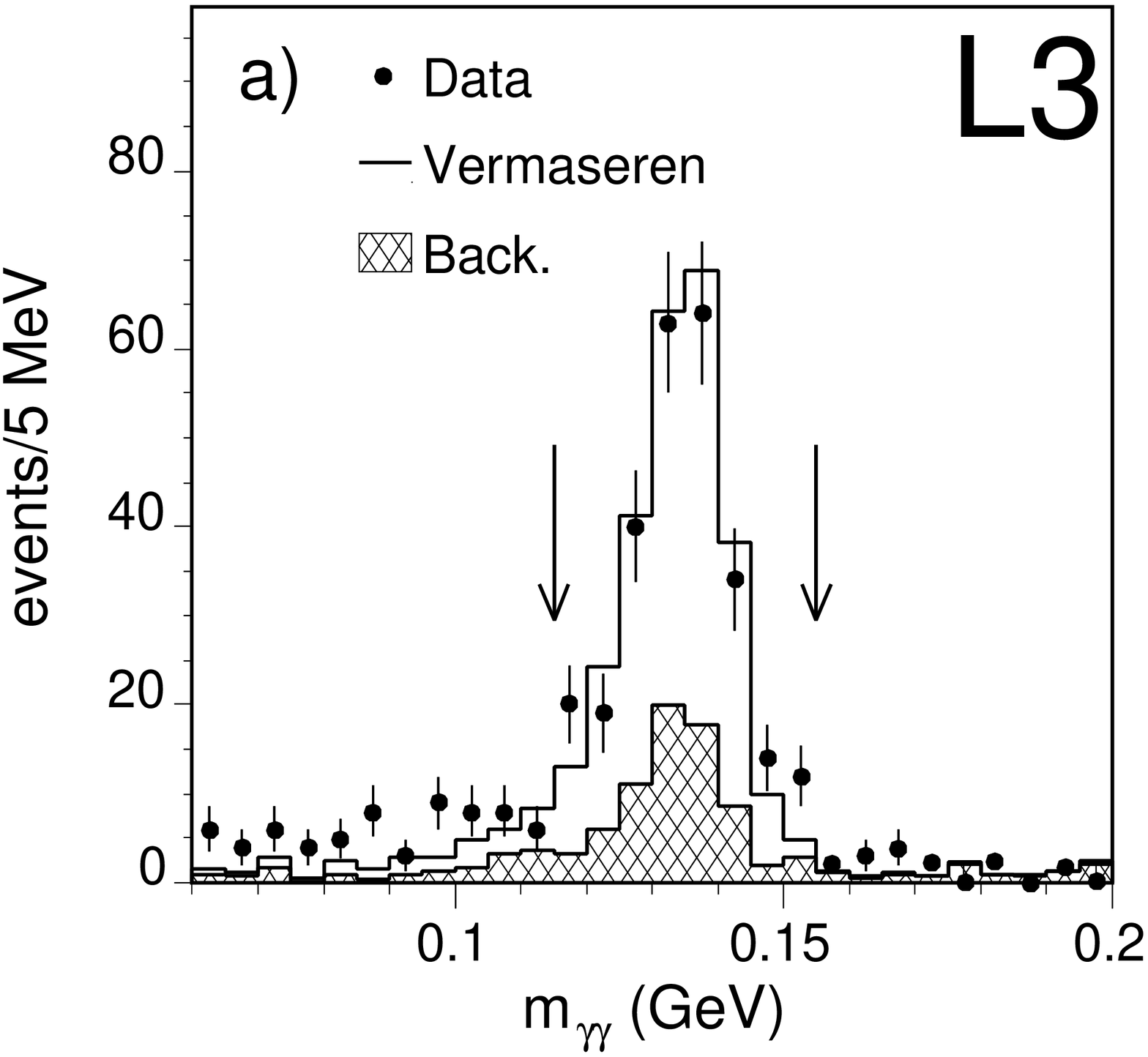,width=82mm,clip=}
\epsfig{file=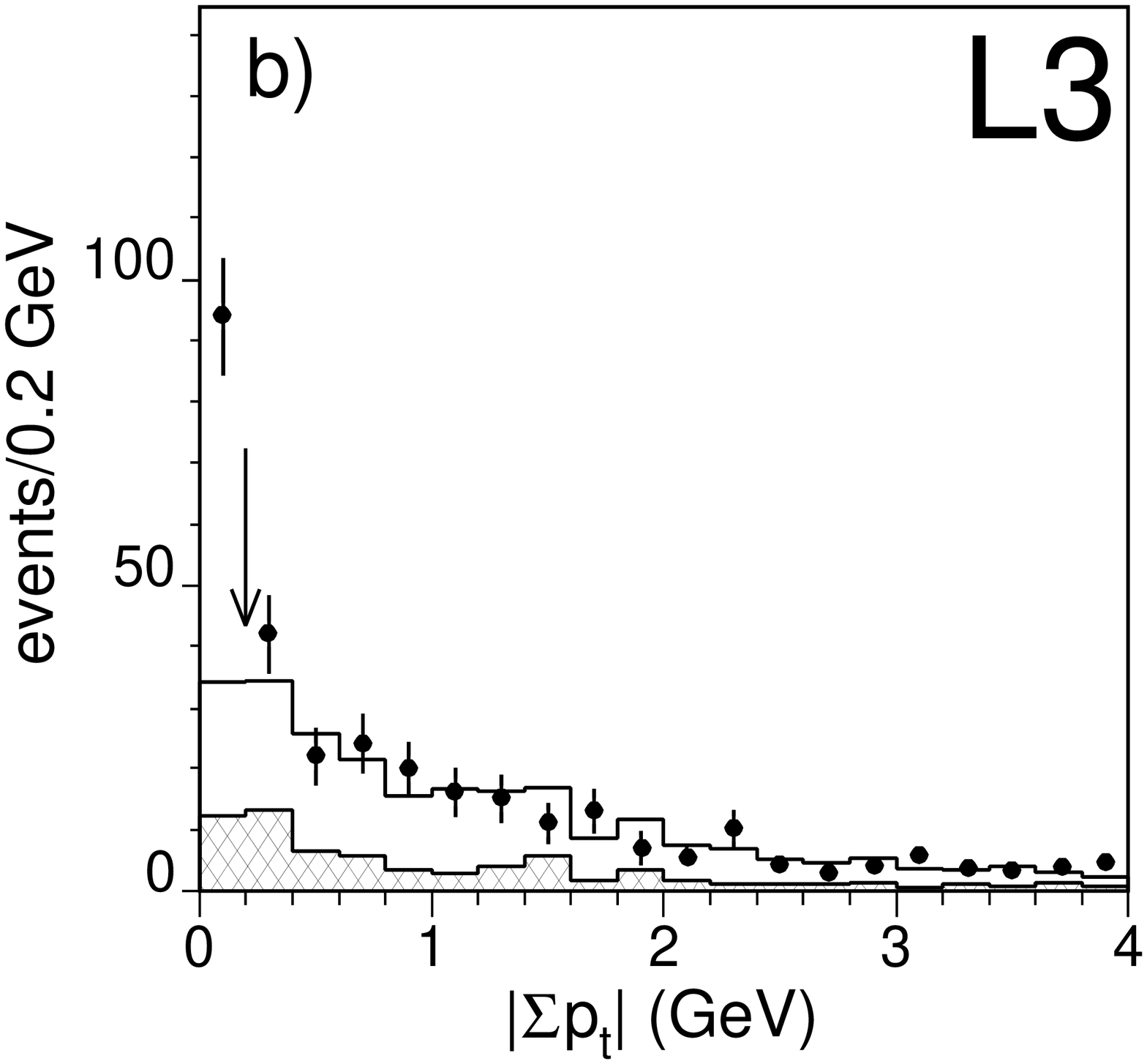,width=82mm,clip=}
\epsfig{file=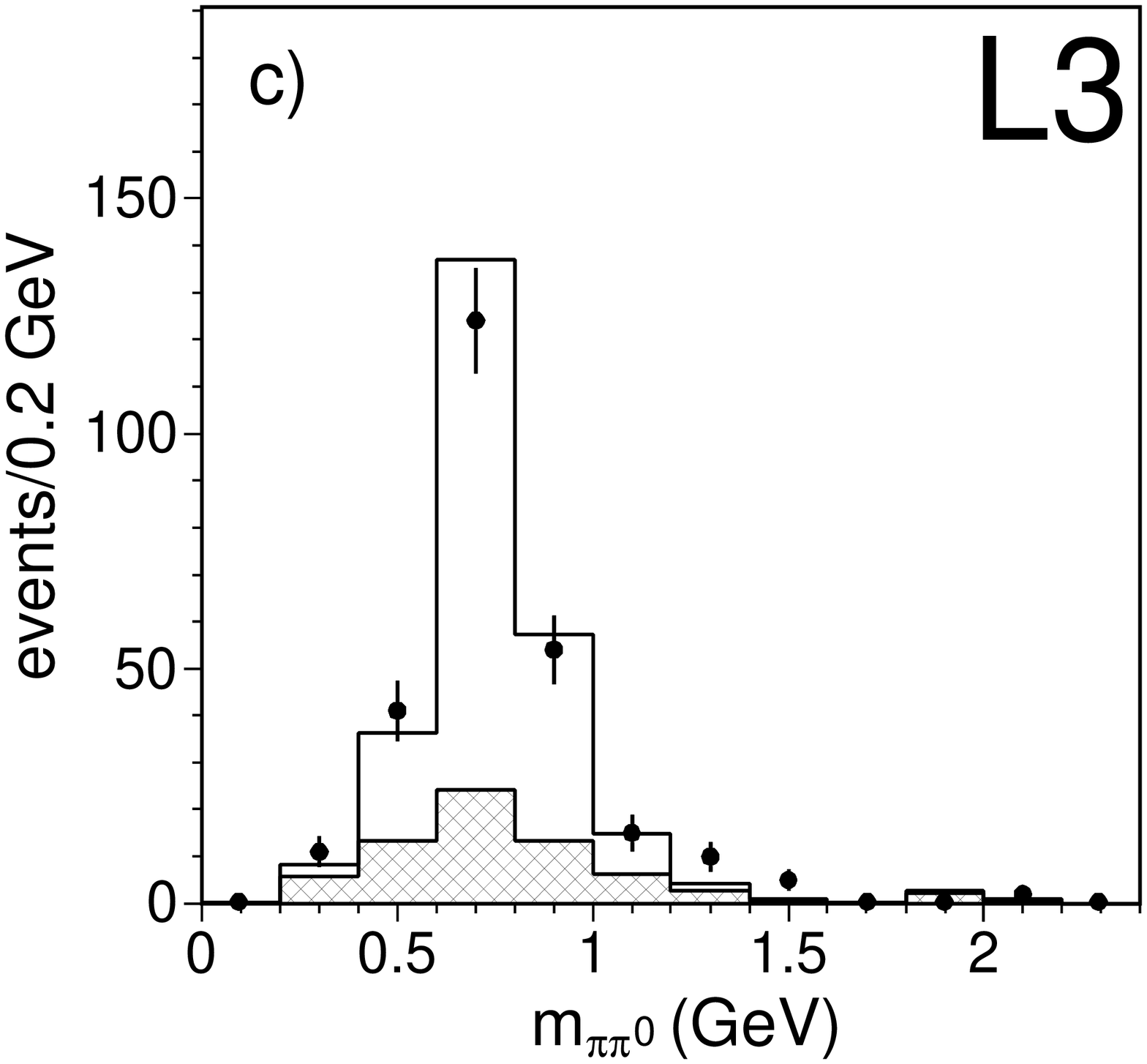,width=82mm,,clip=}
\epsfig{file=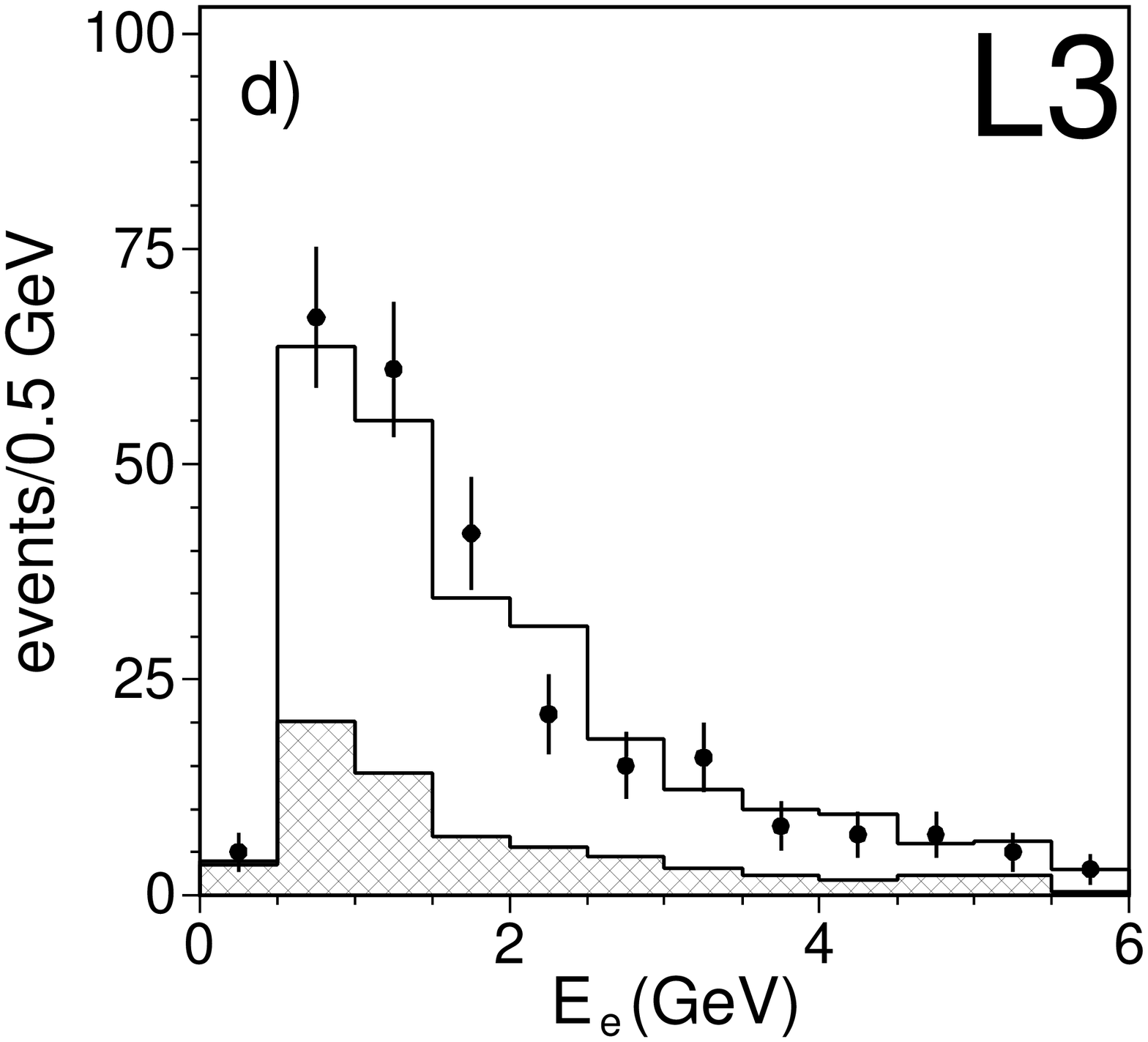,width=82mm,clip=}
\caption{Distributions for $\tau \to \pi\pi\nu$ candidates of a) the
effective mass of final state photons, $m_{\gamma \gamma}$, b) the
sum of the transverse momenta of the charged particles, $|\sum
\vec{p_t}|$, and c) the
effective mass of the two pions, $m_{\pi\pi^0}$. d) Distributions of the energy
of the electron for $\tau \to \e\nu\nu$ candidates.  The data are
compared to the sum of the Vermaseren Monte Carlo $\ee \to \ee\tau^+\tau^-$
and of the background, normalized to the integrated
luminosity. Arrows in a) and b) indicate the position of the cuts on the 
plotted variable, when all other selection cuts are fulfilled}
\label{fig:sele}
\end{center}
\end{figure}

To identify $\tau^+ \rightarrow \pi^+ \pi^0 \bar{\nu_{\tau}}$ decays,
we require the two photons to be compatible with a $\pi^0$.  The
remaining charged particle is considered to be the $\pi^+$
candidate. No additional selection cut is applied on the $\pi^+$. The
two-photon effective mass distribution in Figure~\ref{fig:sele}a shows
the $\pi^0$ peak.  A gaussian fit to this peak gives a mass of
$134.6 \pm 0.6$~\MeV\,and a width of $6.8 \pm 0.7$~\MeV, compatible with the expected
detector resolution.  We require the two-photon effective mass to be
within the range from $115 \MeV$ to $155 \MeV$.  To reject exclusive
final states, as for example $\ee \to \ee \mathrm{a}_2(1320) \to
\ee\pi^+\pi^-\pi^0$, we require the total transverse momentum
imbalance $|\sum \vec{p_{t}}|$ to be greater than 0.2 \GeV.
Figure~\ref{fig:sele}b compares the $|\sum \vec{p_{t}}|$ distribution
of data and Monte Carlo.  The excess of data for $|\sum \vec{p_{t}}| <
0.2\GeV$ is due to exclusive two-photon processes not included in the
Monte Carlo.

With these criteria, 266 events are selected.  As expected for the
$\tau ^+ \to \pi^+ \pi^0 \bar{\nu_{\tau}}$ decay channel, the $\pi^+
\pi^0$ effective mass is consistent with the $\rho$ meson mass, as
shown in Figure~\ref{fig:sele}c.  The energy distribution of the
electron candidate is shown in Figure~\ref{fig:sele}d.  All data
distributions are in good agreement with Monte Carlo simulations.

Table~\ref{tab:selevents} shows the number of observed events together
with selection and trigger efficiencies. The latter are
evaluated directly from the data \cite{thesis}. In the analysis, a two dimensional
trigger efficiency correction, based on the highest momentum track and
the azimuthal opening angle between the two tracks, is applied to each event.
The main background in the sample is
26\% and is due to tau-pairs decaying to other final states, where
leptons or pions are misidentified, or additional pions are not
detected.  The background from the $\ee \rightarrow \tau^+ \tau^-$
process, from resonant final states and from hadron production in
two-photon collisions is less than $4\%$.  The background from
beam-gas and beam-wall interactions is found to be negligible.

\section{Results}

\subsection{\boldmath$\ee \rightarrow \ee \mu^+ \mu^-$}

The cross-section of the process $\ee \to \ee\mu^+\mu^-$ for $3 \GeV <
W_{\gamma\gamma} < 40 \GeV$ is measured for $|\cos\theta_\mu| < 0.8$
and extrapolated to the full angular range. The results are given in
Table~\ref{tab:diagcross} for different values of $\rts$.
 
For lower luminosities the systematic uncertainties are dominated by
the uncertainty on the trigger efficiency, around 3\%. At higher
luminosities the main uncertainty of about 1.5\% arises from the
limited Monte Carlo statistics. The uncertainty due to the event
selection is estimated by varying the selection criteria for the data
samples with high integrated luminosity and is less than 1\%.

\begin{figure}[p]
\begin{center}
\epsfig{file=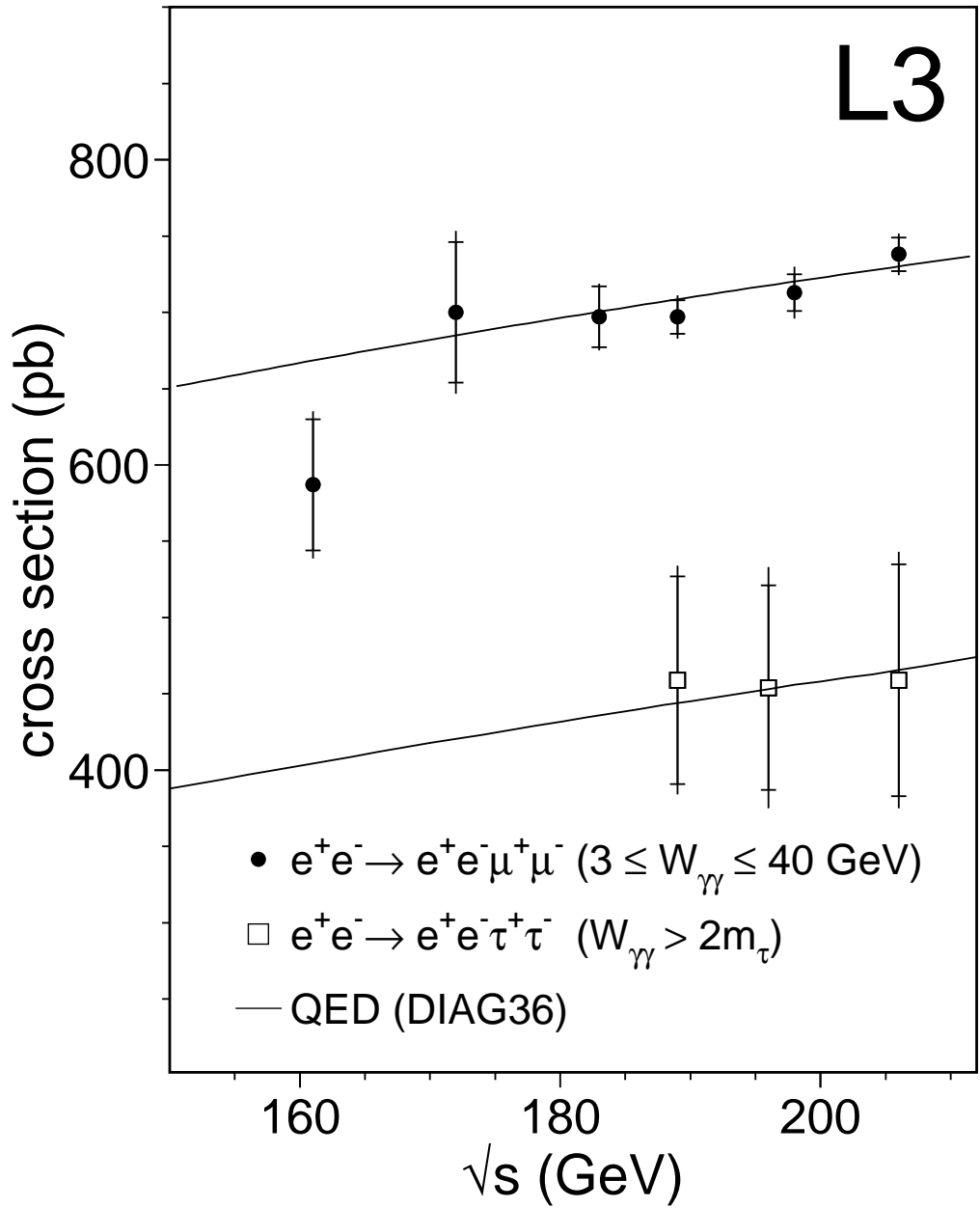,width=140mm,clip=}
\caption{The cross section of the $\ee \to \ee\mu^+\mu^-$ process for
$3 \le W_{\gamma\gamma} \le 40\, \GeV$ and the total cross section of
the $\ee \to \ee\tautau$ process for $W_{\gamma\gamma}>2m_\tau$. The
data are compared to the QED calculations of DIAG36. The inner parts of the error bar represent the statistical uncertainties, 
the outer parts the systematic uncertainties}
\label{fig:cross}
\end{center}
\end{figure}

\begin{figure}[p]
\begin{center}

 \vspace*{-1cm}

\includegraphics*[width=130mm,clip=]{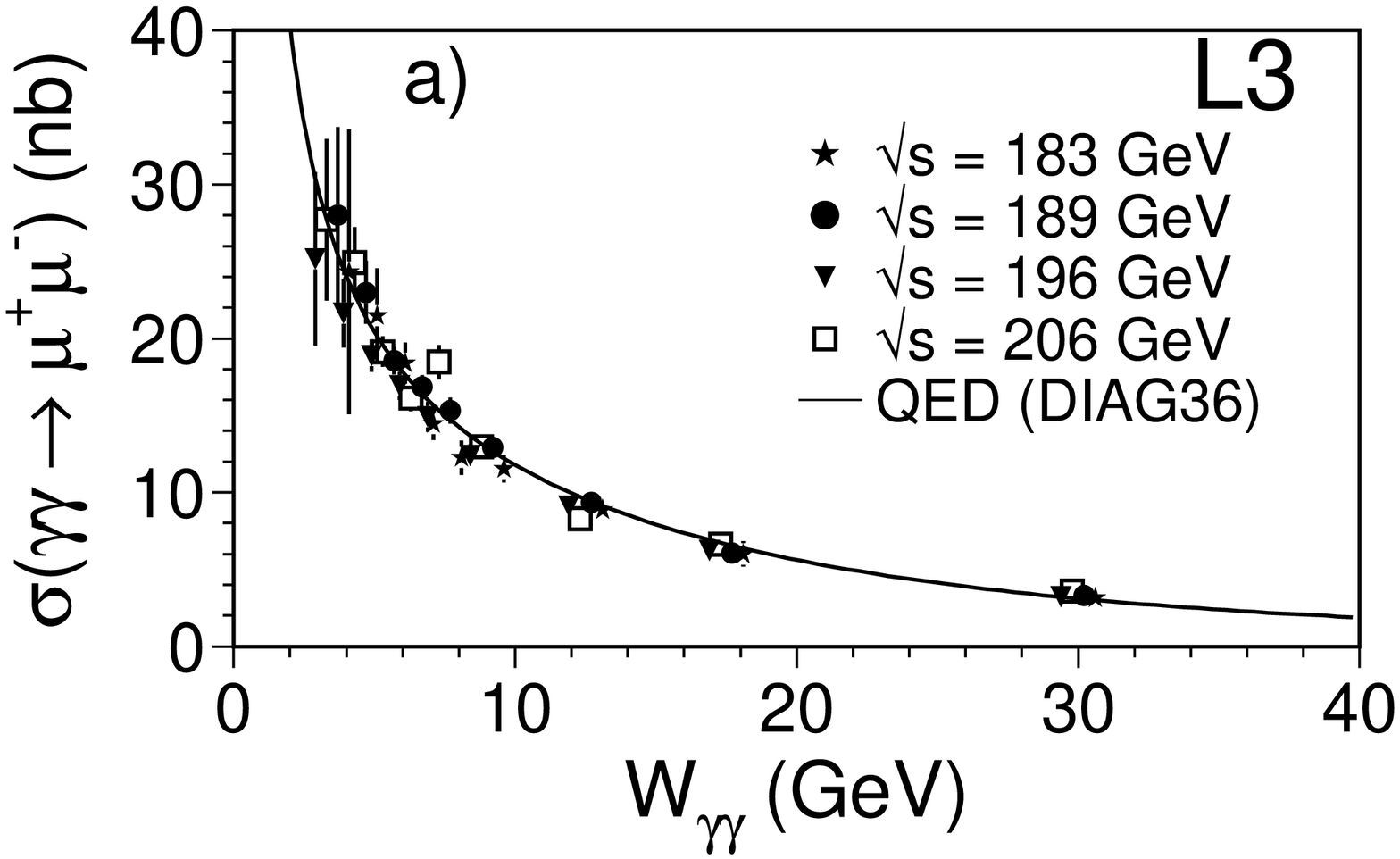}
\includegraphics*[bb=92 701 674 139, width=140mm,clip=]{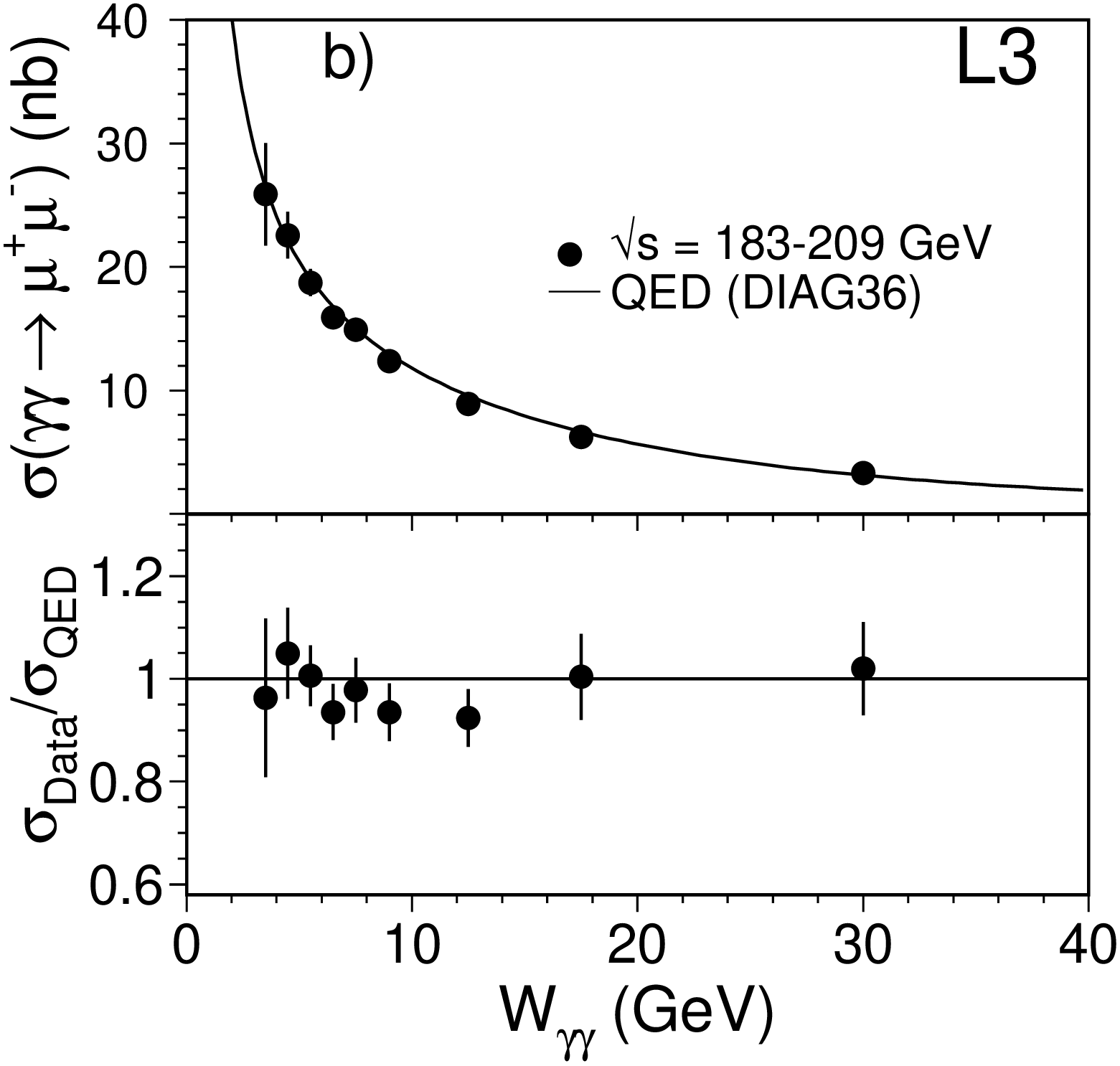}

\vspace*{13cm}
\caption{The cross section of the process $\gamma\gamma \to
\mu^+\mu^-$ as a function of the $\gamma\gamma$ centre-of-mass energy
for a) different values of $\rts$ and b) their combination.  The data
are compared to the QED calculations of DIAG36.}
\label{fig:mures}
\end{center}
\end{figure}

The cross section for the full angular range, presented in
Figure~\ref{fig:cross}, shows the expected slow rise as a function of
$\rts$ and is in good agreement with the QED prediction, as
calculated by DIAG36 Monte Carlo.  The cross section of
the process $\gamma\gamma \rightarrow \mu^+\mu^-$
 is derived by measuring the cross section 
of the $\ee \rightarrow \ee \mu^+ \mu^-$ process   in nine
$W_{\gamma\gamma}$ bins  and scaling it by the two-photon luminosity
function \cite{schuler}. 
The values obtained at different $\rts$
are consistent within a given $W_{\gamma\gamma}$ bin, as shown in
Table~\ref{tab:wgg} and Figure~\ref{fig:mures}a.  Combined results for
the full data sample are listed in Table~\ref{tab:wgg} and shown in
Figure~\ref{fig:mures}b together with the QED predictions. A good
agreement is observed.

\subsection{\boldmath$\ee \rightarrow \ee \tau^+ \tau^-$\unboldmath}

The total $\tau$-pair production cross section is given in
Table~\ref{tab:diagcross}. The
cross section is lower than the $\ee \to \ee\mu^+\mu^-$ cross section
because of the $\tau$-pair mass threshold of 3.6 \GeV.
The main contributions to systematic uncertainties comes from the
variation of the cuts on $|\sum \vec{p_{t}}|$ and the electron
momentum, both between 4\% and 5\%. The total systematic uncertainty
due to selection criteria is estimated to be between 7\% and
9\%. Other sources of systematic uncertainties are the determination
of the trigger efficiency, the Monte Carlo statistics and the
uncertainty on the background level; their combined contribution is
below 3\%. Figure~\ref{fig:cross} compares the measured cross section
and the $\mathcal{O}(\alpha^4)$ QED calculation. A good agreement is
found.

\subsection{Anomalous Couplings of the Tau Lepton}

Photon couplings to the tau lepton are in general due to its
electric charge, the magnetic dipole moment and the electric dipole
moment. They can be described by a matrix element in which the usual
$\gamma^\mu$ term is replaced by~\cite{itzykson}:
\begin{displaymath}
\Gamma^\mu = F_1(q^2)\gamma^\mu +
 iF_2(q^2)\sigma^{\mu\nu}\frac{q_\nu}{2m_\tau} +
 F_3(q^2)\gamma_5\sigma^{\mu\nu}\frac{q_\nu}{2m_\tau},
\end{displaymath}
where the form factors $F_1(q^2)$, $F_2(q^2)$ and $F_3(q^2)$, functions
of the four-momentum squared, $q^2$, of the photon, are related to the tau charge, magnetic
and electric dipole moments as:
\begin{displaymath}
e_\tau = eF_1(0),\,\,\, \mu_\tau={e\left(F_1(0)+F_2(0)\right) \over
  2m_\tau },\,\,\, d_\tau = - { eF_3(0)\over 2m_\tau},
\end{displaymath}
respectively. In the Standard Model, at tree level, $F_1(q^2) = 1$ and
$F_2(q^2) = F_3(q^2) = 0$. Limits on $F_2(q^2)$ and $F_3(q^2)$ were
derived from the decay width $\Gamma(\mathrm{Z \rightarrow
\tau^+\tau^-})$, relating the $\mathrm{Z\tau\tau}$ coupling to the
photon couplings via $SU(2)\times U(1)$
invariance\cite{Escribano:1997wp}. Direct studies of the
$\gamma\tau\tau$ couplings were performed at the Z pole, by the
L3~\cite{Acciarri:1998iv} and OPAL\cite{Ackerstaff:1998mt}
collaborations through the $\rm e^+e^- \rightarrow Z \rightarrow
\tau^+\tau^-\gamma$ process, and at the $\Upsilon(4\rm S)$ by the BELLE collaboration through
the $\rm e^+e^- \rightarrow \gamma^\star \rightarrow \tau^+\tau^-$
process~\cite{Inami:2002ah}.

Tau-pair production in two-photon collisions is sensitive to possible
anomalous couplings of the tau lepton. Values of $F_2(q^2)$ and
$d_\tau$ different from zero would modify the cross section of the
$\rm e^+e^-\rightarrow e^+e^-\tau^+\tau^-$
process~\cite{Cornet:1996pw}. By comparing the measured cross section
with predictions~\cite{Cornet:1996pw} as a function of $F_2(q^2)$ and
$d_\tau$ we obtain:
\begin{displaymath}
|F_2(0)| \le 0.107,\,\,\,\,|d_\tau| \le 1.14 \cdot 10^{-15} ~\mbox{{\it e}\,cm}
\end{displaymath}
at 95\% confidence level, where the limit on each coupling is derived
fixing the other coupling to zero. These bounds, limited by the size
of the data sample, are in agreement with the more stringent published
ones~\cite{Acciarri:1998iv,Ackerstaff:1998mt,Inami:2002ah} and are
derived from a different process.

%
%

\newpage
\typeout{   }     
\typeout{Using author list for paper 281 -  }
\typeout{$Modified: Jul 15 2001 by smele $}
\typeout{!!!!  This should only be used with document option a4p!!!!}
\typeout{   }
%
%
%
%
%
%

\newcount\tutecount  \tutecount=0
\def\tutenum#1{\global\advance\tutecount by 1 \xdef#1{\the\tutecount}}
\def\tute#1{$^{#1}$}
\tutenum\aachen            
\tutenum\nikhef            
\tutenum\mich              
\tutenum\lapp              
\tutenum\basel             
\tutenum\lsu               
\tutenum\beijing           
\tutenum\bologna           
\tutenum\tata              
\tutenum\ne                
\tutenum\bucharest         
\tutenum\budapest          
\tutenum\mit               
\tutenum\panjab            
\tutenum\debrecen          
\tutenum\dublin            
\tutenum\florence          
\tutenum\cern              
\tutenum\wl                
\tutenum\geneva            
\tutenum\hefei             
\tutenum\lausanne          
\tutenum\lyon              
\tutenum\madrid            
\tutenum\florida           
\tutenum\milan             
\tutenum\moscow            
\tutenum\naples            
\tutenum\cyprus            
\tutenum\nymegen           
\tutenum\caltech           
\tutenum\perugia           
\tutenum\peters            
\tutenum\cmu               
\tutenum\potenza           
\tutenum\prince            
\tutenum\riverside         
\tutenum\rome              
\tutenum\salerno           
\tutenum\ucsd              
\tutenum\sofia             
\tutenum\korea             
\tutenum\purdue            
\tutenum\psinst            
\tutenum\zeuthen           
\tutenum\eth               
\tutenum\hamburg           
\tutenum\taiwan            
\tutenum\tsinghua          

{
\parskip=0pt
\noindent
{\bf The L3 Collaboration:}
\ifx\selectfont\undefined
 \baselineskip=10.8pt
 \baselineskip\baselinestretch\baselineskip
 \normalbaselineskip\baselineskip
 \ixpt
\else
 \fontsize{9}{10.8pt}\selectfont
\fi
\medskip
\tolerance=10000
\hbadness=5000
\raggedright
\hsize=162truemm\hoffset=0mm
\def\r{\rlap,}
\noindent

P.Achard\r\tute\geneva\ 
O.Adriani\r\tute{\florence}\ 
M.Aguilar-Benitez\r\tute\madrid\ 
J.Alcaraz\r\tute{\madrid}\ 
G.Alemanni\r\tute\lausanne\
J.Allaby\r\tute\cern\
A.Aloisio\r\tute\naples\ 
M.G.Alviggi\r\tute\naples\
H.Anderhub\r\tute\eth\ 
V.P.Andreev\r\tute{\lsu,\peters}\
% ...
\newpage
\begin{list}{A}{\itemsep=0pt plus 0pt minus 0pt\parsep=0pt plus 0pt minus 0pt
                \topsep=0pt plus 0pt minus 0pt}
\item[\aachen]
 III. Physikalisches Institut, RWTH, D-52056 Aachen, Germany$^{\S}$
\item[\nikhef] National Institute for High Energy Physics, NIKHEF, 
     and University of Amsterdam, NL-1009 DB Amsterdam, The Netherlands
\item[\mich] University of Michigan, Ann Arbor, MI 48109, USA
\item[\lapp] Laboratoire d'Annecy-le-Vieux de Physique des Particules, 
     LAPP,IN2P3-CNRS, BP 110, F-74941 Annecy-le-Vieux CEDEX, France
\item[\basel] Institute of Physics, University of Basel, CH-4056 Basel,
     Switzerland
\item[\lsu] Louisiana State University, Baton Rouge, LA 70803, USA
\item[\beijing] Institute of High Energy Physics, IHEP, 
  100039 Beijing, China$^{\triangle}$ 
\item[\bologna] University of Bologna and INFN-Sezione di Bologna, 
     I-40126 Bologna, Italy
\item[\tata] Tata Institute of Fundamental Research, Mumbai (Bombay) 400 005, India
\item[\ne] Northeastern University, Boston, MA 02115, USA
\item[\bucharest] Institute of Atomic Physics and University of Bucharest,
     R-76900 Bucharest, Romania
\item[\budapest] Central Research Institute for Physics of the 
     Hungarian Academy of Sciences, H-1525 Budapest 114, Hungary$^{\ddag}$
\item[\mit] Massachusetts Institute of Technology, Cambridge, MA 02139, USA
\item[\panjab] Panjab University, Chandigarh 160 014, India.
\item[\debrecen] KLTE-ATOMKI, H-4010 Debrecen, Hungary$^\P$
\item[\dublin] Department of Experimental Physics,
  University College Dublin, Belfield, Dublin 4, Ireland
\item[\florence] INFN Sezione di Firenze and University of Florence, 
     I-50125 Florence, Italy
\item[\cern] European Laboratory for Particle Physics, CERN, 
     CH-1211 Geneva 23, Switzerland
\item[\wl] World Laboratory, FBLJA  Project, CH-1211 Geneva 23, Switzerland
\item[\geneva] University of Geneva, CH-1211 Geneva 4, Switzerland
\item[\hefei] Chinese University of Science and Technology, USTC,
      Hefei, Anhui 230 029, China$^{\triangle}$
\item[\lausanne] University of Lausanne, CH-1015 Lausanne, Switzerland
\item[\lyon] Institut de Physique Nucl\'eaire de Lyon, 
     IN2P3-CNRS,Universit\'e Claude Bernard, 
     F-69622 Villeurbanne, France
\item[\madrid] Centro de Investigaciones Energ{\'e}ticas, 
     Medioambientales y Tecnol\'ogicas, CIEMAT, E-28040 Madrid,
     Spain${\flat}$ 
\item[\florida] Florida Institute of Technology, Melbourne, FL 32901, USA
\item[\milan] INFN-Sezione di Milano, I-20133 Milan, Italy
\item[\moscow] Institute of Theoretical and Experimental Physics, ITEP, 
     Moscow, Russia
\item[\naples] INFN-Sezione di Napoli and University of Naples, 
     I-80125 Naples, Italy
\item[\cyprus] Department of Physics, University of Cyprus,
     Nicosia, Cyprus
\item[\nymegen] University of Nijmegen and NIKHEF, 
     NL-6525 ED Nijmegen, The Netherlands
\item[\caltech] California Institute of Technology, Pasadena, CA 91125, USA
\item[\perugia] INFN-Sezione di Perugia and Universit\`a Degli 
     Studi di Perugia, I-06100 Perugia, Italy   
\item[\peters] Nuclear Physics Institute, St. Petersburg, Russia
\item[\cmu] Carnegie Mellon University, Pittsburgh, PA 15213, USA
\item[\potenza] INFN-Sezione di Napoli and University of Potenza, 
     I-85100 Potenza, Italy
\item[\prince] Princeton University, Princeton, NJ 08544, USA
\item[\riverside] University of Californa, Riverside, CA 92521, USA
\item[\rome] INFN-Sezione di Roma and University of Rome, ``La Sapienza",
     I-00185 Rome, Italy
\item[\salerno] University and INFN, Salerno, I-84100 Salerno, Italy
\item[\ucsd] University of California, San Diego, CA 92093, USA
\item[\sofia] Bulgarian Academy of Sciences, Central Lab.~of 
     Mechatronics and Instrumentation, BU-1113 Sofia, Bulgaria
\item[\korea]  The Center for High Energy Physics, 
     Kyungpook National University, 702-701 Taegu, Republic of Korea
\item[\purdue] Purdue University, West Lafayette, IN 47907, USA
\item[\psinst] Paul Scherrer Institut, PSI, CH-5232 Villigen, Switzerland
\item[\zeuthen] DESY, D-15738 Zeuthen, Germany
\item[\eth] Eidgen\"ossische Technische Hochschule, ETH Z\"urich,
     CH-8093 Z\"urich, Switzerland
\item[\hamburg] University of Hamburg, D-22761 Hamburg, Germany
\item[\taiwan] National Central University, Chung-Li, Taiwan, China
\item[\tsinghua] Department of Physics, National Tsing Hua University,
      Taiwan, China
\item[\S]  Supported by the German Bundesministerium 
        f\"ur Bildung, Wissenschaft, Forschung und Technologie
\item[\ddag] Supported by the Hungarian OTKA fund under contract
numbers T019181, F023259 and T037350.
\item[\P] Also supported by the Hungarian OTKA fund under contract
  number T026178.
\item[$\flat$] Supported also by the Comisi\'on Interministerial de Ciencia y 
        Tecnolog{\'\i}a.
\item[$\sharp$] Also supported by CONICET and Universidad Nacional de La Plata,
        CC 67, 1900 La Plata, Argentina.
\item[$\triangle$] Supported by the National Natural Science
  Foundation of China.
\end{list}
}
\vfill


\newpage


\begin{table}[p]
\begin{center}
\begin{tabular}{|c|c|r|r@{ $\pm$ }r|r@{ $\pm$ }r|r|r|}
\hline
 & $\langle\rts\rangle$ [$\GeV$]  & \multicolumn{1}{c|}{$\int \mathcal{L} \rm{d}t~$ [pb]} & \multicolumn{2}{c|}{$\epsilon_{\ell}~[\%]$} & \multicolumn{2}{c|}{$\epsilon_{\rm trig}~[\%]$} & 
\multicolumn{1}{c|}{$N_{D}$} & \multicolumn{1}{c|}{$N_{B}$} \\ \hline 
$\mu^+\mu^-$	& 161	&  $10.2$\hspace{4mm} &$18.4$ & $ 0.5$	    &$99.4$ & $ 0.6$  & 193  &4\hspace{2mm} 	\\
		& 172	&  $9.7$\hspace{4mm} &$18.9$ & $ 0.5$        &$98.4$ & $ 0.8$  & 223  &7\hspace{2mm}		\\
		& 183	&  $54.2$\hspace{4mm} & $18.4$ & $ 0.3$    &$99.7$ & $ 0.2$  & 1188 & 15\hspace{2mm} \\
		& 189	& $170.3$\hspace{4mm} & $20.1$ & $ 0.3$    &$99.6$ & $ 0.1$  & 4025 & 33\hspace{2mm} \\
		& 196	& $154.0$\hspace{4mm} & $18.9$ & $ 0.3$    &$99.7$ & $ 0.1$  & 3491 & 36\hspace{2mm} \\
		& 206	& $192.7$\hspace{4mm} & $19.1$ & $ 0.2$    &$99.7$ & $ 0.1$  & 4576 & 45\hspace{2mm} \\
\hline
$\tau^+\tau^-$	& 189   & $172.1$\hspace{4mm} &  $1.18$ & $ 0.04$  &$71.8$ & $ 1.3$  &$85$ & 25\hspace{2mm} \\
		& 196   & $220.9$\hspace{4mm} &  $1.29$ & $ 0.05$  &$60.1$ & $ 1.6$  &$97$ & 31\hspace{2mm} \\
		& 206   & $215.1$\hspace{4mm} &  $1.08$ & $ 0.04$  &$58.0$ & $ 0.9$  &$84$ & 29\hspace{2mm} \\
\hline
\end{tabular}
\caption{Centre-of-mass energies and corresponding integrated
luminosities.  The selection efficiency, $\epsilon_\ell$, and trigger
efficiency, $\epsilon_{\rm trig}$, are also given together with the number of
observed events, $N_D$, and the background contribution, $N_B$.
}
\label{tab:selevents}
\end{center}
\end{table}
\begin{table}[p]
\begin{center}
\begin{tabular}{|c|c|r@{ $\pm$ }r@{ $\pm$ }r|c|r@{ $\pm$ }r@{ $\pm$ }r|c|}
 \hline
 & $\langle\rts\rangle$ [$\GeV$] & \multicolumn{3}{c|}{$\sigma_{\rm{DATA}}$ [pb]}  & $\sigma_{\rm{QED}}$ [pb]  & \multicolumn{3}{c|}{$\sigma_{\rm{DATA}}$ [pb]} & $\sigma_{\rm{QED}}$ [pb]  \\
& & \multicolumn{3}{c|}{$|\cos\theta_\mu|<0.8$} & $|\cos\theta_\mu|<0.8$ &\multicolumn{3}{c|}{} & \\
 \hline
$\mu^+\mu^-$    & 161  & $101.4 $ & $ 7.2 $ & $ 2.6 $& 115.4 &$587$ & $43$ & $22$&  668.3\\
                & 172  & $119.2 $ & $ 7.6 $ & $ 3.1 $& 116.6 &$700$ & $46$ & $27$&  684.9\\
                & 183  & $117.7 $ & $ 3.4 $ & $ 1.9 $& 118.3 &$697$ & $20$ & $9$ &  700.7\\
                & 189  & $117.1 $ & $ 1.8 $ & $ 1.8 $& 118.9 &$697$ & $11$ & $9$ &  708.6\\
                & 196  & $118.9 $ & $ 2.0 $ & $ 2.2 $& 120.3 &$713$ & $12$ & $12$&  717.8\\ 
                & 206  & $122.6 $ & $ 1.8 $ & $ 1.7 $& 121.3 &$738$ & $11$ & $8$ &  730.0\\ \hline
 $\tau^+\tau^-$ & 189  &\multicolumn{3}{c|}{}&& $459$ & $68$ & $33$&  442.6 \\
                & 196  &\multicolumn{3}{c|}{}&& $454$ & $67$ & $42$&  452.3 \\
                & 206  &\multicolumn{3}{c|}{}&& $459$ & $76$ & $35$&  466.0\\

\hline
\end{tabular}
\caption {The cross sections of the processes $\ee\rightarrow
\ee\mu^+\mu^-$ and $\ee\rightarrow \ee\tau^+\tau^-$ with their
statistical and systematic uncertainties at different $\rts$ values
compared to QED \protect\cite{BDK} expectations. The cross section for
$\ee \rightarrow \ee\mu^+\mu^-$  for $3 \GeV < W_{\gamma\gamma} < 40
\GeV$ is given for both $|\cos\theta_\mu|<0.8$
and for the full solid angle.}
\label{tab:diagcross}
\end{center}
\end{table}

\begin{table}[p]
\begin{center}
\begin{tabular}{|r@{ -- }r|c|c|c|c|c|c|}
\hline
\multicolumn{2}{|c|}{} & \multicolumn{6}{c|}{$\sigma(\gamma\gamma \to \mu^+\mu^-)$ [nb]}\\ \cline{3-8}
\multicolumn{2}{|c|}{\raisebox{1.5ex}[-1.5ex]{$W_{\gamma\gamma}$}} & & & & & & \\
\multicolumn{2}{|c|}{\raisebox{1.5ex}[-1.5ex]{[\GeV]}} & {\raisebox{1.5ex}[-1.5ex]{$\rts=$ 183
\GeV}} & {\raisebox{1.5ex}[-1.5ex]{189 \GeV}} & {\raisebox{1.5ex}[-1.5ex]{196
\GeV}} & {\raisebox{1.5ex}[-1.5ex]{206 \GeV}} &
{\raisebox{1.5ex}[-1.5ex]{$183-209 \GeV$}}&{\raisebox{1.5ex}[-1.5ex]{QED}}\\
\hline
3  & 4    & $ 24.3\pm 9.9 $& $ 28.0\pm 6.6 $ & $ 25.2\pm 6.4 $ & $ 27.7\pm 6.1 $ & $ 25.9\pm 4.2 $ & $26.8$\\
4  & 5    & $ 21.5\pm 3.7 $& $ 23.0\pm 2.7 $ & $ 25.2\pm 2.8 $ & $ 24.9\pm 3.0 $ & $ 22.6\pm 1.9 $ & $21.5$\\\
5  & 6    & $ 18.4\pm 1.9 $& $ 18.6\pm 1.5 $ & $ 21.6\pm 1.6 $ & $ 19.1\pm 1.6 $ & $ 18.7\pm 1.1 $ & $18.6$\\
6  & 7    & $ 14.5\pm 1.5 $& $ 16.8\pm 1.3 $ & $ 18.8\pm 1.4 $ & $ 16.1\pm 1.3 $ & $ 15.9\pm 0.9 $ & $17.0$\\
7  & 8    & $ 12.3\pm 1.5 $& $ 15.3\pm 1.3 $ & $ 14.9\pm 1.4 $ & $ 18.5\pm 1.7 $ & $ 14.9\pm 1.0 $ & $15.2$\\
8  & 10   & $ 11.5\pm 1.3 $& $ 12.9\pm 1.0 $ & $ 12.4\pm 1.1 $ & $ 12.9\pm 1.1 $ & $ 12.4\pm 0.7 $ & $13.2$\\
10  & 15  & $ \phantom{0}8.9\pm 1.0 $& $  \phantom{0}9.3\pm 0.8 $ & $  \phantom{0}9.1\pm 0.8 $ & $  \phantom{0}8.3\pm 0.7 $ & $  \phantom{0}8.9\pm 0.5 $ & $ 9.6$\\
15  & 20  & $ \phantom{0}6.0\pm 1.0 $& $  \phantom{0}6.1\pm 0.7 $ & $  \phantom{0}6.2\pm 0.8 $ & $  \phantom{0}6.6\pm 0.8 $ & $  \phantom{0}6.2\pm 0.5 $ & $ 6.2$\\
20  & 40  & $ \phantom{0}3.1\pm 0.6 $& $  \phantom{0}3.3\pm 0.4 $ & $  \phantom{0}3.2\pm 0.5 $ & $  \phantom{0}3.6\pm 0.5 $ & $  \phantom{0}3.3\pm 0.3 $ & $ 3.2$\\

\hline
\end{tabular}
\caption {The cross section of the process $\gamma\gamma \to
\mu^+\mu^-$ with its combined statistical and systematic uncertainties
as a function of $W_{\gamma\gamma}$ for four different $\rts$ values and
their average together with the QED \protect\cite{BDK} expectations.}
\label{tab:wgg}
\end{center}
\end{table}




\end{document}